
%
%
%
%
%



\def \Mpch {h^{-1}{\rm Mpc}}
\def \etal {{\it et al.\/}}

\magnification=1200
\baselineskip .5truecm
\tolerance=10000
\hsize 17.0truecm
\vsize 25.0truecm
\voffset 1.4truecm

\nopagenumbers
\def\ref{\par\noindent\hangafter=1\hangindent=1truecm}
\def\parn{\par\noindent}

{\centerline{\bf{MAPPING THE DARK MATTER IN CLUSTERS}}}
\vskip 1.2truecm
{\centerline{\bf{ Nick Kaiser$^{1,2}$, Gordon Squires$^1$}}}
{\centerline{\bf{ Greg Fahlman$^3$ and David Woods$^3$}}}
\vskip 0.8truecm
$^1$ CITA, University of Toronto, Canada  \par
$^2$ Canadian Institute for Advanced Research Cosmology Programme  \par
$^3$ Astronomy and Geophysics, UBC, Vancouver, Canada

\vskip 10.0truecm
\noindent
{\bf ABSTRACT} \smallskip
Massive clusters of galaxies gravitationally
shear the images of faint background
galaxies. At large impact parameters
the shear is weak, but can still be measured,
to a reasonable degree of significance, as a statistical anisotropy
of the faint galaxy images.  We describe techniques for measuring the
shear and discuss the interpretation of the shear field.  We have applied
this analysis to ms1224+007. We find a clear detection of the shear,
but, puzzlingly, we find a mass about three times that obtained
by application of the virial theorem, and obtain a very large mass-to-light
ratio.  Similar results have been obtained by other groups, and we
discuss their implications.
\par

\vfill\eject
\baselineskip .6truecm
\voffset 0.0truecm

Giant arcs in clusters provide a very clean probe of the mass distribution
in the central parts of clusters (see e.g.~the reviews of Soucail
and Miralda-Escude in these proceedings).
The work we shall describe here follows the pioneering
study of A1689 by Tyson, Valdes and Wenk (1990) where they measured
for the first time the statistical anisotropy of the background galaxies
at much larger radii.

\bigskip\noindent
{\bf{ MEASURING THE IMAGE SHEAR}} \smallskip
{}From deep photometry of a field containing a massive cluster we first
find the faint galaxies and measure their central second angular
moments: $Q_{ij} = \int d^2 \theta W(\theta) \theta_i \theta_j f(\vec \theta)$,
where $f$ is the surface brightness,
and form the ``polarization''
$$
e_\alpha = \left[\matrix{
(Q_{11}-Q_{22}) / (Q_{11}+Q_{22})\cr
2 Q_{12} / (Q_{11}+Q_{22})\cr
}\right]
$$
The window function $W(\theta)$ we use is a gaussian with scale length matched
to that of the galaxy.  In the absence of lensing the $e_\alpha$
values will scatter around a mean of zero
on the $e_1,e_2$ plane, but a gravitational shear
acting coherently over some angle will displace the distribution.
Provided the shear is weak --- and we will
be measuring shears of typically 10\% --- the shift in the mean polarization
is $\langle e_\alpha \rangle = P_{\rm sh} s_\alpha$,  where the shear $s_\alpha
\equiv \{\phi_{,11} - \phi_{,22}, 2\phi_{,12}\}$ and the surface
potential $\phi$ satisfies $\nabla^2\phi = 2 \sigma$ with dimensionless
surface density $\sigma = \Sigma_{\rm phys} / \Sigma_{\rm crit}$.
The effective critical surface density is $\Sigma_{\rm crit}^{-1}
= 4\pi G a_l w_l \langle \max(0, 1 - w_l/w_g) \rangle$ with the average being
taken
over the distribution of comoving distances $w_g$ to the faint galaxies.
The proportionality constant $P_{\rm sh}$
--- which we call the shear polarizability --- depends on the details of the
shapes of the galaxies, and can be estimated either for the population
as a whole
or, as we do, individually for each galaxy.

Each galaxy therefore provides an estimate of the shear along a particular
line of sight, though a rather noisy one
as there is a substantial scatter in the intrinsic polarizations.  These can
be averaged together to produce e.g.~a smoothed shear map $s(\vec \theta)$
with statistical uncertainty $\sim \sqrt{\langle s^2 \rangle_{\rm
intrinsic}/N}$
where  $N$ is the number of galaxies being averaged.
An interesting feature of cluster studies is that the signal to noise
varies rather slowly with radius; at large radii the signal becomes small,
but the number of galaxies goes up and these cancel if the surface density
varies as $\Sigma \propto 1/r$ for instance.

A minor complication is that distortion of the images can arise in the
telescope and in the earth's atmosphere.  Luckily there are sufficiently
many foreground stars to provide a control sample with which to measure
the point spread function quite precisely.  One could imagine reconvolving the
image with
a psf designed to recircularise the stars and this would then null
out the systematic error in the galaxies.  In fact what we do
is to calculate for each galaxy a `smear polarizability' $P_{\rm sm}$ analogous
to
$P_{\rm sh}$ which tells us how the polarization shifts in response
to smearing by
an anisotropic psf ($P_{\rm sm}$ is essentially a measure of the inverse
area of the galaxy), and use this to null out the systematic error.  This
approach is easier if, as in the data we describe later, the psf anisotropy
varies across the chip.  An ultimate limit on this technique
--- which one might eventually hope to apply to very large fields where it
is vital to cancel even very tiny systematic effect --- may possibly
arise due to effects such as atmospheric dispersion and chromatic aberrations
of the optics which cause the psf to depend on the spectrum of the
objects, but with present data such effects are small compared to
the statistical error.

\bigskip\noindent
{\bf{ INTERPRETATION OF THE SHEAR }} \smallskip

 Assuming we have a map of the shear $s(\vec \theta)$ how do
we infer from this the surface density $\sigma$?  First, there is no
local relation between the shear and $\sigma$.  Physically, this reflects the
fact that
a constant density sheet lens does not produce shear.  As one might expect,
however,
there is a local relation between the angular gradients of $s$ and of $\sigma$:
$$
\matrix{
{\partial \sigma / \partial x} = {1\over 2}
\left[{\partial s_1 / \partial x} + {\partial s_2 /\partial y}\right]
\cr
{\partial \sigma / \partial y} = {1\over 2}
\left[{\partial s_2 /\partial x} - {\partial s_1 / \partial y}\right]
}
$$
This means that in principle one can construct any {\sl differential\/}
measurement
of the surface density; for instance one can calculate the surface density
difference between two points simply by integrating $\nabla \sigma$ along
some line: $\delta \sigma_{12} =
\sigma(\vec\theta_1) - \sigma(\vec\theta_2) = \int_{\theta_1}^{\theta_2}
\vec{dl}
\cdot \nabla \sigma$.  The arbitrariness of the line chosen reflects the
inherent non-uniqueness of any $\sigma$ determination method.  This arises
essentially because one has two inputs $s_1,s_2$ from which we only
want to recover the single scalar function $\sigma$.  In the example here
one could estimate $\delta \sigma_{12}$ by averaging over any combination
of paths from $\theta_1$ to $\theta_2$ --- and one would presumably try to find
some optimum weighted
combination of these --- and one could also use loop integrals in some way to
provide a check on the quality of the data, though how best to do this has
yet to be worked out in any systematic way.

The ambiguity in the baseline surface density is something of a problem,
particularly when the data coverage is limited.  The ambiguity
can be resolved to some extent by studying clusters with giant arcs,
though this involves some uncertain interpolation of the surface density
from the radii where the arcs lie out to the radii where the weak shear
analysis
can be safely applied.  An alternative is to try and measure the perturbation
of the background galaxy counts $n(m)$ (Broadhurst, Peacock and Taylor,
1994) to determine the surface density, though this is
quite difficult, as we show below.
We have developed a number of tests derived from the relation above between
$\partial \sigma / \partial \theta_i$ and $\partial s_\alpha / \partial
\theta_i$:

\smallskip\noindent
{\bf{ Mass Imaging }} \smallskip

Kaiser and Squires (1992, hereafter KS93)
provide an algorithm to reconstruct a smoothed
2-dimensional mass image.  In the present context, this can be viewed
as an average over all radial paths from infinity (or some large radius
where the surface density and the shear can be assumed small) to the
point in question.  This becomes a two dimensional integral
$\hat \sigma = \langle \int dl \cdot \nabla \sigma \rangle_{\rm radial\ paths}
\rightarrow \int d^2 \theta \ldots$ which in turn can be replaced by a sum
over the background galaxies (provided these are sufficiently dense on the
sky).

An important practical limitation of this method comes from boundary
terms introduced when the data are finite.  Near the centre of the
mass reconstruction this causes a constant negative shift which can be
expressed
in terms of the surface density and shear on the boundary (for a circular
field the shift is the mean of $\sigma$ plus half the mean tangential
shear around the boundary).  Further out nearer the edge of the field one finds
a spurious negative trough; the lack of observed shear beyond the
field effectively introduces an integral constraint which forces the total
mass in the lens to be zero.
A nice feature of the method is that while the
estimator is a convolution of the observed shears with an extended kernel,
the noise has a white spectrum and, provided we average over a scale containing
several galaxies, should be quite accurately gaussian.
This makes assigning significance to features in the mass
reconstructions fairly straightforward.

A weakness of the method is that it does not appear to make full use of the
redundancy in the data described above.  We have developed an alternative
method which creates the most probable density field compatible with the
data, under the prior assumption of gaussian noise with user specified
colour and amplitude.  This should incorporate the extra information.  The
results
appear quite similar in {\sl shape\/}
to those obtained with the KS93 method, but there is a
bias in amplitude in the reconstructed  $\sigma(\vec \theta)$ which is hard to
calibrate.

\smallskip\noindent
{\bf{ Laplacian Map}} \smallskip

An alternative is to construct a map of the laplacian of $\sigma$:
$$
\nabla^2 \sigma = D_\alpha s_\alpha
$$
where $D_\alpha \equiv \{\partial^2/\partial x^2-\partial^2/\partial y^2,
2 \partial^2 / \partial x \partial y\}$.   As we want a smoothed map of
the laplacian we implement  this as a filter in fourier space,
the $D_\alpha$ operator
becoming an algebraic function which multiplies the
smoothing filter transfer function.
This relation is local, so we don't need to worry about boundary terms.
This method is probably going to be most useful
for attempting to find clusters blindly in future large survey fields; a
cluster
will appear as a negative dip.  A nice feature of this method is that one
can also use the redundancy in the input data to calculate a map of
$\nabla \times \nabla \sigma$ --- which is simply obtained by applying the
`rotation' $s_1 \rightarrow s_2,\ s_2 \rightarrow -s_1$ to the data before
applying
the $D_\alpha$ operator --- and which should of course be zero.
This provides a useful check on the consistency of the data.

\smallskip\noindent
{\bf{ Aperture Massometry}} \smallskip

It is possible to put a rigorous lower bound on the mass contained
within a circular aperture.
This method uses the mean tangential
shear around a circular path of radius $\theta$:
$$
\langle s_T \rangle =
\int {d\phi \over 2 \pi} \left[ s_1 \cos 2 \phi + s_2 \sin 2 \phi\right]
= -{d \overline \sigma \over d \ln \theta}
$$
where $\overline \sigma$ is the mean surface density within the circle
and the second equality (which is trivial for a circularly symmetric lens)
follows in the general case from the 2-dimensional version of Gauss' law.
It is then easy to show that the statistic
$$
\zeta(\theta_1, \theta_2) \equiv
(1 - \theta_1^2 / \theta_2^2)^{-1}
\int\limits_{\theta_1}^{\theta_2} d\ln \theta \langle s_T \rangle
= \overline \sigma(<\theta_1)- \overline \sigma(\theta_1 <
\theta <\theta_2)
$$
where the last symbol represents the mean surface density in the annulus
$\theta_1 < \theta < \theta_2$.  Since this is necessarily non-negative
$\zeta$ provides a lower bound on $\overline \sigma(<\theta_1)$.  The
$\int d\ln \theta \langle s_T \rangle$ can again be replaced by an
area integral and thereby by a discrete sum over galaxies and, as
with the massmap determination, the error analysis is straightforward.

An interesting feature of this analysis is that it uses only data which
lie in the control annulus, so provided a sufficiently large field, this
can be placed outside of most of the cluster light, greatly simplifying
the problem of distinguishing background galaxies from cluster members.

Unless the ratio of inner and outer radii $\theta_2 / \theta_1$ is made very
large, in which case we will only be able
to estimate $\overline \sigma$ in a very small region,
the systematic underestimation of $\overline \sigma$ can be quite substantial.
For a power law surface density profile $\sigma \propto \theta^{-\gamma}$
and with $\theta_2 = a \theta_1$, $\zeta / \overline \sigma =
(a^2 - a^{2-\gamma}) / (a^2 - 1)$, so for $\gamma = 1$, $\zeta$ underestimates
the true $\overline \sigma$ by 33\% and 25\% for $a = 2, 3$ respectively.

As with the laplacian map, it is possible to obtain a check on the
consistency of the data by `rotating' the inputs,
$s_1 \rightarrow s_2,\ s_2 \rightarrow -s_1$ which should result
in $\zeta = 0$ within the statistical error.

The $\zeta$ statistic is really a special case of the KS93 method for
a particular choice of smoothing kernel; in this case a `compensated
top-hat'.  In fact, from $d \overline \sigma / d \ln r =
- \langle s_T \rangle$ one can show that an estimator for the
surface density smoothed with an arbitrary circular
window function with zero total weight: $\int d^2r  W(r) = 0$ is
$$
\int d^2 r W(r) \sigma(\vec r) = 2 \pi \int dr
\langle s_T \rangle W'(r)
$$
where
$$
W'(r) = {1\over r} \int_0^r dr' r' W(r') - {r W(r) \over 2}
$$
which provides a simple way to construct the window function
$W'$ for e.g.~any mexican hat type $W(r)$.
Note that if $W$ vanishes beyond some radius then so does $W'$,
which again is nice if one is dealing with finite data.

\smallskip\noindent
{\bf{ Dilution of the Counts}} \smallskip

The ambiguity in the baseline surface density is a nuisance.
This can be resolved in principle by measuring the dilution
of the background counts $\delta n(m)$ caused by the
amplification (BPT).
Under the weak lens assumption, the perturbation to the counts
is just proportional to the surface density, though
with a rather small constant of proportionality since the
amplification bias for faint galaxies is rather small.  There are clearly
some technical difficulties in applying this method; here one
must look for a perturbation {\sl under\/} the lens, so to speak, rather
than around it as in the shear measurement, so one must be
careful to correct for the faint cluster galaxy counts.  This can
be aided by using colour information,  but getting accurate colours
is expensive.  One would also have to correct for masking of background
galaxies by the high surface brightness parts of foreground galaxies, and
perhaps subtle effects in the
image detection and photometry where the extended diffuse
light around the cluster galaxies overlays the background.

With effort these problems can perhaps be overcome, but the
following example suggests that the method will
still suffer from rather low signal to noise:
If we use the $\zeta$ statistic to estimate
$\overline\sigma$, the statistical uncertainty is
$\langle \hat {\overline \sigma} \rangle^{1/2} =
(1-1/a^2)^{-1} \sqrt{
\langle s_1^2 \rangle / (4\pi \overline n r_1^2)
}$ where $r_2 = a r_1$, $\overline n$ is the surface number
density of galaxies,
and the rms shear noise (with both instrumental and intrinsic
contributions) is
$\langle s_1^2 \rangle^{1/2} \simeq 0.43$ for the
data described below.
{}From the observed faint galaxy counts, the
amplification bias appears to be quite small; $\simeq -0.25$.
The amplification is $2 \overline \sigma$, so the
corresponding estimator is
$\hat {\overline \sigma} \simeq - 2 \delta n(m) / n(m)$ (BPT describe
alternatives such as looking for a change in the slope of the
counts, but the simple example here illustrates the basic idea and
it is hard to imagine that the noise in any other estimator is
likely to be significantly different).  Assuming a poisson
distribution for the background galaxies, the
 noise in this estimator is then $\langle \hat {\overline \sigma}
^2\rangle^{1/2} =
2 / \sqrt{\pi \overline n r_1^2}$. Clustering
of the background galaxies will inflate this
by an uncertain but probably appreciable factor, but the minimal
poisson error is already
about a factor 8 larger
than that for the $\zeta$ statistic, so for the ms1224 data described
below, for example, the expected S/N is below unity.

\bigskip\noindent
{\bf{ MS1224 }} \smallskip

This cluster was chosen for its high X-ray luminosity, its high
redshift ($z = 0.33$) making it possible to survey a $\sim 2 \Mpch$ square
field in a reasonable time, and because it was also a target of the
CNOC cluster project.  The optical spectroscopy studies
(Carlberg, Yee and Ellingson, 1994, hereafter CYE) gave
a modest velocity dispersion of $\simeq 750$km/s, and apparently consistent
with this, a low richness, though as we shall see, the mass found from
the lensing appears to be much greater.

As described in Fahlman \etal, 1994 (hereafter FKSW),
we took 1 hr total I-band integrations
on each of four fields surrounding the cluster centre under excellent seeing
conditions.
Our software found $\sim 5000$ objects over an area (after allowing
for masked regions around bright foreground stars) of about 120 square arcmin,
from which we extracted a `faint-extended' subsample of about  2000 galaxies
covering a range of about 3 magnitudes to I=23.4.
As a test,
these data were artificially stretched, rebinned onto a $2\times$ coarser pixel
grid and then
degraded to simulate the effects of seeing. These synthetic
data were then analysed
in the same way as  the real data.  This verified
that the analysis software was indeed able to detect the artificial
shear,
and also allowed us to estimate a small, but critical `signal loss factor'
due to seeing and other biases in the analysis.
 The anisotropy of the psf was
measured and the correction applied as described above.

In the real data a clear shear signal was seen: the mass-map has a peak which
coincides
quite well with the bulk of the smoothed cluster light measured by
CYE, and the tangential shear was also clearly seen (as a significance
level of about 5-sigma).  The $\zeta$-statistic gave
$\overline \sigma(<2.7') \ge 0.06 \pm 0.012$.
The signal was also seen repeatedly in independent
magnitude sub-samples.  A mysterious dark peak also appears repeatedly
in the mass-map. This is not seen in the cluster light, but as it is only
a $\simeq$3-sigma detection it should not be taken too seriously.

\noindent
Fig. 1.---Spatial distribution and polarization (or ellipticity)
parameters for the ms1224 faint galaxy subsample.  The left panel
shows the individual galaxies to be fairly uniformly distributed
on the sky,
though with some holes around bright foreground stars and galaxies.
No particularly strong coherent distortion is apparent to the eye.
However, in the panel on the right we have smoothed the polarizations
to make a map of the shear, and the characteristic shear pattern
is now clearly seen.

While the {\sl location\/} of the main mass peak is in nice accord with
the optical measurement, the {\sl amplitude\/} is not.  The
mass scale calibration depends on the effective inverse critical
surface density which in turn depends on the redshift distribution
of the background galaxies.  For our brighter galaxies we can
measure $\Sigma_{\rm crit}^{-1}$ directly using the data
of Lilly, 93 and Tresse \etal, 1993; the much larger CFRS redshift survey
has recently been completed, increasing the sample size by
an order of magnitude, and seems to reinforce the results
obtained from the smaller published samples.  For the fainter galaxies
we need to extrapolate.  Extrapolating the trend at brighter magnitudes
suggest a very slow increase in $\Sigma_{\rm crit}$ of about 30\%
per magnitude.  A similar or even more modest increase is indicated
by the slow variation of shear with magnitude limit seen in our
studies of ms1224 and now of several other similar clusters.

The upshot of all of this is a lower bound (aside from statistical error, of
course)
on the mass within a radius of $2.76'$ (or roughly $0.5\Mpch$
in physical radius) of
$M_{\rm lens} \ge 3.5 \times 10^{14} M_\odot/h$.  This can be compared with
the virial mass in the same aperture --- obtained by taking the total
virial mass estimate and multiplying by the fraction of the total
cluster light lying within the aperture --- of
$M_{\rm virial} = 1.15 \times 10^{14} M_\odot/h$; a factor 3 smaller
than $M_{\rm lens}$.  We can also use our photometry (or that of CYE)
to obtain a mass to light ratio $(M/L)_V \simeq 800 h$, much larger than
the values typically obtained for nearby optically selected
clusters.

\noindent
Fig. 2.---Contour plot of the projected mass reconstruction superposed
on the summed I-band image of the central $7'$ square region of
ms1224.
\vfill\eject

We have thought long and hard about possible errors or biases in our
analysis which might have corrupted our mass estimates.
Here is a non-exhaustive list of factors we have considered, though most
of these can be safely discounted:
{\it i) Biases:\/}
There are a number of
biases: $\zeta < \overline \sigma$; the optical light estimate includes
projected
material in the centre beam, but does not cover the entire control aperture,
and no attempt has been made to remove the cluster galaxies, which will
tend to dilute the shear signal.  Allowing for these biases
would only exacerbate the discrepancy.
{\it ii) Statistical Uncertainty:\/}
The statistical uncertainty in $\overline \sigma$ is about 20\%.  At the time
we submitted FKSW there was a larger statistical uncertainty arising from
the rather small size of the redshift surveys, but the larger CFRS survey
seems to give similar results.  The redshift surveys are, of course, not
100\% complete and the missing 10\% or so of
the galaxies are most
probably at higher redshift.  However,  even if we put these at redshift
2 say, the effect on our mass estimate is small.  We found that in
order to reconcile the mass estimates would require a median galaxy redshift
$\sim 4$, which is clearly unreasonable.
There is some variation in the redshift distribution from field to field in
the $z$-surveys, and while the fluctuations expected for the larger field
here are smaller, there is a remote
possibility that we hit an extreme statistical
fluctuation in $n(z)$
due to large-scale structure. However, it is relevant to note that the {\sl
counts\/}
of galaxies we find agree well with those of e.g.~Lilly,
Cowie and Gardner (1991).
{\it iii) Contamination:\/}  Our method  measures the total projected mass
(excess) within our aperture.  It is possible (though extremely improbable)
that there is another high mass object within the aperture.  However, if this
lies at a redshift $>0.33$ then the projected
mass-to-light ratio for the combined system only
increases, and if it lies at lower redshift it is hard to
see why it was not detected.  If we had no knowledge at all of the lens
redshift,
the lowest mass to light ratio is obtained by  placing the lens at a somewhat
lower redshift, and is only marginally less than our quoted value.
We consider the
effect of adding mass around the cluster itself below.
{\it iv) Inhomogeneous Sampling:\/}  Our massmap algorithm assumes uniformly
sampled data, whereas in fact there are some masked holes which can result
in a bias in the shape of real observed signal (though they do not create a
bias in the sense of causing spurious
detections where there is no real signal).  Could
this have created the dark-blob?
We have explored this with
simulations obtained by using the real galaxy positions but by shuffling
the shear estimates and adding a known lensing signal, and we find any
such bias to be very small.
{\it v) Weak Shear Assumption:\/} If the shear becomes strong then the assumed
linear response of the population mean polarisation will fail. In particular
one might worry that
gravitational amplification might have increased the mean redshift of the
background galaxies behind the cluster.  However, the measured shear in ms1224
is
in  fact very weak: $s\sim 10\%$ or so after correction for seeing; the light
in
the cluster is compact (more than half the light lies within our central
aperture)
so if mass traces light at least, the amplification for the galaxies we
actually
use is very small and in any case all indications are that the median
redshift increases very weakly with increasing magnitude.  One might worry
that even though the mean surface density is low, the dark matter has
clumpy substructure giving localised regions of non-linearity
which somehow bias the mean shear estimation.  This possibility probably
deserves detailed study, but one argument against this is that one often sees
very long smooth arcs, suggesting the dark matter in clusters is actually
smoothly distributed, and,
as emphasised by Tyson (1990), if one makes the dark mass too
clumpy this produces a large number of conspicuous arcs which are
not seen.
{\it vi) Cosmological Model:\/}  Our calculation of $\Sigma_{\rm crit}$
assumes a flat, zero cosmological constant model.  Changing these assumptions
makes only a very slight change to our results.
{\it vii) Correlated intrinsic ellipticities:\/}
Our error analysis assumes that the intrinsic
ellipticities are uncorrelated.  Flin (1993a,b) finds a tendency
for galaxies in physical pairs and triplets to be aligned, but this is a weak
statistical effect seen at the $\simeq 2$-sigma level in a sample of
$\sim 10^3$ galaxies, and would negligibly effect the noise level
in our mass-maps. We are currently analysing some blank fields where we can
check for
spurious detections.  None have been found as yet.
{\it viii) Signal loss factor:\/}
Our simulations, which show that due to seeing etc.~we recover
only about 70\% of the
input shear, assume that the faint galaxies are simply scaled down replicas of
their
brighter cousins.  This may be false, but the sizes of the
synthetic and real galaxies at least are quite similar.  A better
way to establish this calibration factor would be stretch and then degrade
deep images from HST, which will shortly become possible.  We are confident
however, that seeing does not {\sl increase\/} the shear.

None of these loopholes appear very promising and it is therefore
hard to escape the conclusion that, in this cluster at least, the mass
and the mass-to-light ratio are indeed very large.  We have subsequently
obtained similarly extensive data on A2218, and smaller fields
on A2163 and A2390.  These clusters all have arcs with
measured redshifts, and while our analysis is
still ongoing, we see no indication that the weak shear analysis
overestimates the mass as compared with the arcs.

\vfill\eject
\noindent
Fig. 2.---Contour plot of the projected mass reconstruction superposed
on the photometry in A2218.
\vfill\eject

\bigskip\noindent
{\bf{ WHAT DOES IT MEAN?}} \smallskip

There are really two puzzles here: Why is $M_{\rm lens} \gg M_{\rm virial}$
and why is the lensing derived $M/L$ so large.  It is quite possible to imagine
that the observed velocity dispersion underestimates the mass: after all, the
well documented `beta-discrepancy' problem seems to indicate a sizeable
scatter in $\sigma_v^2 / T_X$ and we might just be seeing a low-beta
cluster.  Perhaps the cluster consists of two clumps merging along a
direction perpendicular to the line of sight and that is why we see a low
$\sigma_v$.
It is interesting that the conspicuous giant elliptical in the cluster
does not lie near the
centre of the main cluster concentration, reinforcing the suspicion that we
are not dealing with a well relaxed system.  This might
explain a low virial mass, but would not reduce
the $M/L$.

Another possibility is that the galaxies are relaxed, but that they suffer
from velocity bias because their scale length is shorter than that of the
mass.  The key question is what is the mass profile derived from the shear?
Unfortunately this is rather difficult to answer as with the current data
we only have a 5-sigma detection, so there is considerable noise as well
as bias in the mass profile. In addition, one should be wary of
dilution of the shear signal
by cluster galaxies which will further bias the profile
in the centre.  Our lower bound
on the
projected aperture mass applies if the projected mass vanishes in the control
annulus:
i.e.~implicitly assuming a very {\sl steep\/} mass profile.
The lensing data themselves would be quite compatible with
a more extended mass distribution.
Under the empty annulus assumption the shear should fall
off within the annulus as $s\propto 1/r^2$ whereas in fact it appears to
be remarkably flat with radius, suggesting, at face value, a very flat surface
density profile indeed.
However, we must emphasise that invoking an extended mass profile will {\sl
increase\/}  the mass to light ratio if one self-consistently corrects
for the bias in the $\zeta$ statistic.
With an isothermal sphere type mass
profile, for example, $\overline \sigma$, and therefore
$M/L$ {\sl within our aperture\/}  would increase by about 30\%, and the
$M/L$ within $1 \Mpch$
would increase even more.  Nor is it clear
that this type of solution can really explain the low observed velocity
dispersion; in the isothermal sphere example, the 1-D velocity
dispersion for any population of finite radial extent is
$\sigma_v^2 = V^2_{\rm rot}/ 3 = \theta \overline \sigma(<\theta) /
(6 \pi) = (940 {\rm km/s})^2$, still larger than that
observed.

Even our minimal mass gives a surprisingly high mass-to-light ratio
(or essentially equivalent, but easier to measure, a very high mass
per galaxy). However, the value is not at all out of line with other
studies using the same technique: Bonnet \etal, 1993 claim the large
scale shear around cl0024 requires a mass roughly 3-times larger than
the virial mass, and Smail \etal, 1994 quote $M/L$ of $\simeq 550h$,
from their studies of cl1455, cl0016,
quite comparable to the value here.  What makes these high $M/L$ ratios
more surprising is when we allow for the considerable {\sl increase\/} in the
comoving number density of $\sim L_*$
galaxies inferred from the faint galaxy redshift surveys.  We can clearly
measure
the excess cluster counts $N_c$ in our central aperture over about a two
magnitude
range below $L_*$, and thereby obtain $(M/N)_{\rm cluster}$.
Similarly,
we can readily estimate the comoving number density
of field galaxies over the same magnitude range at this redshift and thereby
obtain a estimate of $(M/N)$ for a closed universe.
An estimate
of $\Omega$ then follows if one
assumes that the mass-per-galaxy in the cluster is representative of the
universe as a whole:
$$
\Omega = {(M/N)_{\rm cluster} \over (M/N)_{\rm universe}}
=
{\overline \sigma d\Omega \sqrt{1+z_c} (dn/dz)_{z_c} \over
3 N_c w_l \langle \max(0, 1 - w_l / w_g) \rangle} \simeq 1.8
$$
This is very large compared to the typical values found applying the
same kind of analysis to low redshift, optically selected clusters.
The difference stems roughly equally from the high cluster $M/L$
and from the evolution of the field galaxy population.
There is of course considerable uncertainty in this estimate
due to the uncertain $n(z)$ for the background galaxies, and, as with
any $\Omega$ estimate of this kind, one is really under no obligation
to believe that the mass-per-galaxy of this particular
cluster, or indeed of clusters in general, is representative.

How do we understand the high $M/L$'s if they are indeed real?
Does the mass-to-light ratio of a cluster decrease with time; implying either
that galaxy rich matter falls in later or that somehow galaxy formation
is stimulated within the cluster?
Or perhaps is the explanation simply that there is really a
wide variation in $M/L$'s for clusters; the well studied optically selected
clusters
at  low redshift preferentially seeing the low $M/L$ cases and the arcs
selected clusters naturally sampling the higher end.  This is attractive,
but it is not easy to see why ms1224 should have been biased in this way; the
highly speculative possibility that ms1224 got into the EMSS sample by
macro-lensing a background AGN will shortly be testable with ROSAT and
ASCA spectra.

How could a strong variation in $M/L$ on cluster scales arise?  Could
it be that in an early stage of explosive galaxy formation the bulk of the gas
was
disturbed in the manner envisaged by Ostriker and Cowie
(1981), resulting
in a highly inhomogeneous gas entropy and density distribution?  The evolution
of the dark-matter clustering would proceed essentially undisturbed, but
there will be DM concentrations which happen to lie in regions of high entropy
gas where galaxy formation might plausibly have been impeded.  An appeal of
this idea
is that this might also help explain the `baryon catastrophe' problem
(White, \etal, 1993).
The dark-clumps seen in some of the mass reconstructions are certainly
of interest in this regard, but the significance of the dark feature in the
ms1224 map at least is only marginal.

The results described here and those of Tyson's group, the Toulouse
group and the Durham/Caltech group clearly show that these
observations are a practical way to directly map the dark matter
in clusters.  The high mass-to-light ratios obtained are admittedly somewhat
puzzling.  The calibration of this method, on which these results rest, is
not perfect, but we have argued that it is very hard to make the discrepancy
with the virial mass in ms1224 go away and indeed, at face value,
the lensing data are quite compatible with an even larger mass.  The great
strength of this
method is that it makes no assumptions regarding the shape, dynamical
stability or state of relaxation of the cluster.  There are still uncertainties
in e.g the
redshift distribution of the faint galaxies, which affects the calibration,
but these are small and of an `engineering'
nature and should be solvable with a combination of HST imaging, ground-based
spectroscopy to fainter limits
and a much larger sample of weak-shear
cluster studies.

These observations  are currently limited by detector technology. It would be
of
great value to obtain data out to large radii.  CFHT has
a corrected field $\sim50'$  across, yet we currently use only use a $7'$
square
2048$^2$ chip.  The 4096$^2$ MOCAM array will speed observations by a factor
4.  With a thinned mosaic of this size on a 10m telescope the ms1224 study
could be made in about 4-minutes, so it would be quite practical to go much
fainter, increasing the number of background galaxies substantially, and
thereby
boosting the precision of the measurement.  A further boost in signal  to noise
can be obtained by studying clusters at somewhat lower redshift, though to
explore
the same physical radius becomes more expensive and one also
becomes more sensitive to
how well correlated psf variations and other systematic effects can be
corrected for.  With these developments it should be quite feasible to obtain
detailed individual mass profiles and shapes for the most massive clusters
and also, with large random field surveys, to obtain a mass-selected sample
of clusters.  Galaxies, groups and poor clusters will be hard to
detect individually, but by stacking results it should be possible to
determine e.g.~the galaxy-mass cross-correlation function directly.
Another window of opportunity is to study coherent shear on the scale
of superclusters; current observations being right at the level of precision
where we expect to see a signal appearing.  This will potentially provide a
direct
measure of the mass power spectrum $P_\rho(k)$, giving a strong test of
cosmogonical theories and, combining with COBE type measurements on a
similar scale, giving us a handle on the relative contribution of
tensor and scalar modes and/or the ionisation history.
Finally, a further
spin-off from these studies will be quite accurate measurement of the
mean relative geometrical distances to faint galaxies as a function of
their
size and magnitude.  Combining these with directly measured redshifts should
allow
a fundamental test of the cosmological world model.

\bigskip\noindent
{\bf{ REFERENCES }} \smallskip
\noindent
Tyson, J., Valdes, F., and Wenk, R., 1990. ApJ, 349, L19 \parn
Broadhurst, T., Peacock, J., and Taylor. 1994, preprint \parn
Kaiser, N., and Squires, G., 1993. ApJ, 404, 441 (KS93) \parn
Fahlman, G., Kaiser, N., Squires, G., and Woods, D., 1994.
to appear in ApJ. Available by anonymous ftp from
{\tt ftp.cita.utoronto.ca} in {\tt /cita/nick/ms1224/ms1224.ps}\parn
Carlberg, R., Yee, H. and Ellingson, E., 1994. To appear in ApJ\parn
Lilly, S., Cowie, L, and Gardner. 1991. ApJ, 369, 79 \parn
Flin, P., 1993a. AJ, 105, 473\parn
Flin, P., 1993b, ApJ, 406, 395 \parn
Ostriker, J. and Cowie, L., 1981. ApJ Lett, 285, L127\parn
Bonnet, H., Fort, B., Kneib, J-P., Mellier, Y., and Soucail, G.,
1993. A\&A, 280, L7 \parn
Smail, I., Ellis, R., Fitchett, M. \& Edge, A., 1994. Preprint \parn
Tyson, J., 1990. In proceedings of AIP meeting
``After the First Three Minutes'', eds
Holt, Bennet and Trimble\parn
Tresse, L., Hammer, F., le Fevre, O., and Proust, D., 1993
A\&A, 277, 53\parn
Lilly, S.J., 1993. ApJ, 411, 501\parn
White, S., Navarro, J., Evrard, G., and Frenk, C.,
1993. Nature, 366, 429\parn
\vfill\eject
\end